\documentclass[aps,prd,twocolumn,floatfix,nofootinbib,superscriptaddress]{revtex4-1}
\usepackage{dcolumn}

\usepackage{amsmath,amsfonts,amssymb,mathrsfs,times,bm}
\usepackage{extarrows}
\usepackage{graphicx}
\usepackage{float}
\usepackage{changes}
\usepackage{graphicx,epstopdf}
\usepackage{dcolumn}
\usepackage{exscale}
\usepackage{relsize}
\usepackage[colorlinks=true,breaklinks=true,linkcolor=blue,citecolor=blue,urlcolor=blue]{hyperref}
\newcommand{\nn}{\nonumber}

\begin{document}

\title{Equivalence of Gibbons-Werner method to geodesics method in the study of gravitational lensing}
\author{Zonghai Li}
\affiliation{School of Physical Science and Technology, Southwest Jiaotong University, Chengdu 610031, China}
\affiliation{Center for Theoretical Physics, School of Physics and Technology, Wuhan University, Wuhan 430072, China}
\author{Tao Zhou}
\email{taozhou@swjtu.edu.cn}
\affiliation{School of Physical Science and Technology, Southwest Jiaotong University, Chengdu 610031, China}

\date{\today}

\begin{abstract}
The Gibbons-Werner method where the Gauss-Bonnet theorem is applied to study the gravitational deflection angle has received much attention recently. In this paper, we study the equivalence of the Gibbons-Werner method to the standard geodesics method, and it is shown that the geodesics method can be derived with the Gibbons-Werner method, for asymptotically flat case. In the geodesics method, the gravitational deflection angle of particle depends entirely on the geodesic curvature of the particle ray in the Euclidean space. The gravitational deflection of light in Kerr-Newman spacetime is calculated by different technologies under the Gibbons-Werner framework, as an intuitive example to show the equivalence.
\end{abstract}


\maketitle

\section{Introduction}
Gravitational lensing plays an important role in gravitational theory. In theoretical physics, it is used to test the fundamental theory of gravity, where a famous example is that Eddington\textit{et al.}~\cite{DED1920,Will2015} verified Einstein's general relativity by means of the deflection experiment of light in the solar gravitational field 100 years ago. In astrophysics and cosmology, it is used to measure the mass of galaxies and clusters~\cite{Hoekstra2013,Brouwer2018,Bellagamba2019}, and to detect dark matter and dark energy~\cite{Vanderveld2012,cao2012,zhanghe2017,Huterer2018,SC2019}. In mathematics, it is related to singularity theory, topology and Finsler geometry~\cite{GW2008,Gibbons2009-1,Gibbons2009-2,Caponio2016,Werner2012}.

Recently, Gibbons and Werner~\cite{GW2008} introduced an elegant geometrical method of deriving the bending angle of light in a static and spherically symmetric spacetime. They used the famous Gauss-Bonnet (GB) theorem to a surface defined by the corresponding optical metric. Later, Werner~\cite{Werner2012} extended this method to the rotating and stationary spacetimes. In stationary spacetimes, the optical geometry is defined by the Randers-Finsler metric. Thus, Werner applied Naz{\i}m's method to construct an osculating Riemannian manifold where one can easily use the GB theorem. The work by Gibbons and Werner promotes the study of light deflection. On one hand, Jusufi \textit{et al.}~\cite{Jus-161,Jus-171,Jus-172,Jus-173,Jus-174,Jus-175,Jus-181,Jus-182,Jus-183,Jus-184,Jus-185,Jus-186,Jus-187,Jus-191,Jus-a1,zhu2019} studied the gravitational lensing not only in asymptotically flat spacetime but also in nonasymptotically flat spacetime such as a spacetime with cosmic string. Similar works can also be found in Refs.~\cite{SO-B2,OSS2018,Arakida2018,OV2018,Goulart2018,Javed2019,AO2019,Leon2019}. On the other hand, Ishihara \textit{et al.}~\cite{ISOA2016,IOA2017-1,IOA2017-2,IOA2018,IOA2019} studied the finite-distance corrections for gravitational deflection of light both for the weak and the strong deflection limit, where the source and observer are no longer assumed to be infinitely far apart from a lens. For a review on finite-distance corrections, we refer the reader to Ref.~\cite{OA2019}.

It is well known that there are many massive particles in our Universe, such as massive neutrinos. The study of gravitational deflection of massive particles allows one to understand the properties of the sources and these particles. In fact, the study of the massive particles lensing using traditional methods can be found in Refs.~\cite{AR2002,AP2004,Bhadra2007,Yu2014,He2016,He2017a,He2017b,Jia2016,Jia2019}. Moreover, two other routes have been established by applying the GB theorem to study the gravitational deflection of massive particles. The first route is related to the Jacobi metric of curved spacetime. To be precise, one can calculate the deflection angle of massive particles via applying the GB theorem to the surface defined by the Jacobi metric~\cite{Gibbons2016,LHZ2019} for static spacetime and by the Jacobi-Maupertuis Randers-Finsler metric~\cite{Gibbons2019} for stationary spacetime. The second route is related to the optical media method. For static and spherically symmetric spacetime, Crisnejo and Gallo~\cite{CG2018} used the GB theorem to study the gravitational deflections of light in a plasma medium and the deflection angle of massive particles. The finite-distance corrections of light with a plasma medium and the gravitational deflection of charged massive particles were studied quite recently~\cite{CG2019,CGV2019}. For rotating and stationary spacetimes, Jusufi~\cite{Jus-massive1} used the GB theorem to study the deflection angles of massive particles by the Kerr black hole and the Teo wormhole, respectively, based on the corresponding isotropic type metrics, the refractive index of the corresponding optical media. Furthermore, the method in Ref.~\cite{Jus-massive1} was extended to distinguish naked singularities and Kerr-like wormholes~\cite{Jusufi2019-1}, and to study the gravitational deflection of charged particles in Kerr-Newman spacetime~\cite{Jusufi2019-2}.

In this paper, the method with the GB theorem to study the deflection angle shall be called the Gibbons-Werner method. It is worth investigating whether the Gibbons-Werner method~\cite{GW2008} is equivalent to the standard geodesics method~\cite{Weinberg1972}. In fact, this topic has been discussed by some researchers. The first-order equivalence has been shown in Refs.~\cite{Jus-172,Jus-173,Jusufi2019-1,Jus-a1}, and the second-order equivalence has been shown in Refs.~\cite{CG2018,LHZ2019}. From a conceptual point of view, however, the two methods seem to be completely different. The Gibbons-Werner method shows that the deflection of particles (photon and massive particles) is determined by a quantity outside of itself relative to the lens~\cite{Werner2012,Jus-massive1}, and thus the gravitational deflection angle can be regarded as a global topological effect, whereas the geodesics method is usually associated within a region from particles ray to lens. In the present paper, we will demonstrate the equivalence between the Gibbons-Werner method and the geodesics method for asymptotically flat spacetime, in terms of results and concepts. More specifically, the weak gravitational deflection of light in Kerr-Newman spacetime will be taken as a simple example.

This paper is organized as follows. In Sec.~\ref{TF}, we review the GB theorem and use the theorem to the lens geometry. Then, we show that the equivalence of the Gibbons-Werner method to geodesics method. In Sec.~\ref{exam}, we give the Kerr-Newman spacetime as an example to show the equivalence. Finally, we summarize our results in Sec.~\ref{CONCLU}. Throughout this paper, we use the natural units where $G = c = 1$ and the metric signature $(-,+,+,+)$.
\section{The Equivalence between the Gibbons-Werner Method and Geodesics Method} \label{TF}

\subsection{The Gauss-Bonnet theorem}
Let $D$ be a compact oriented two-dimensional Riemannian manifold with the Euler characteristic $\chi(D)$ and Gaussian curvature $K$, and its boundary $\partial{D}$ is a piecewise smooth curve with geodesic curvature $k_g$. Then, the GB theorem states that~\cite{GW2008,Carmo1976}:
\begin{equation}
\iint_D{K}dS+\oint_{\partial{D}}k_g~d\sigma+\sum_{i=1}{\theta_i}=2\pi\chi(D),\\
\end{equation}
where $dS$ is the area element of the surface, $d\sigma$ is the line element along $\partial{D}$, and $\theta_i$ is the exterior angle defined for the $i$th vertex in the positive sense.

\subsection{Application the Gauss-Bonnet theorem to the lens geometry}
\begin{figure}[t]
\centering
\includegraphics[width=8.5cm]{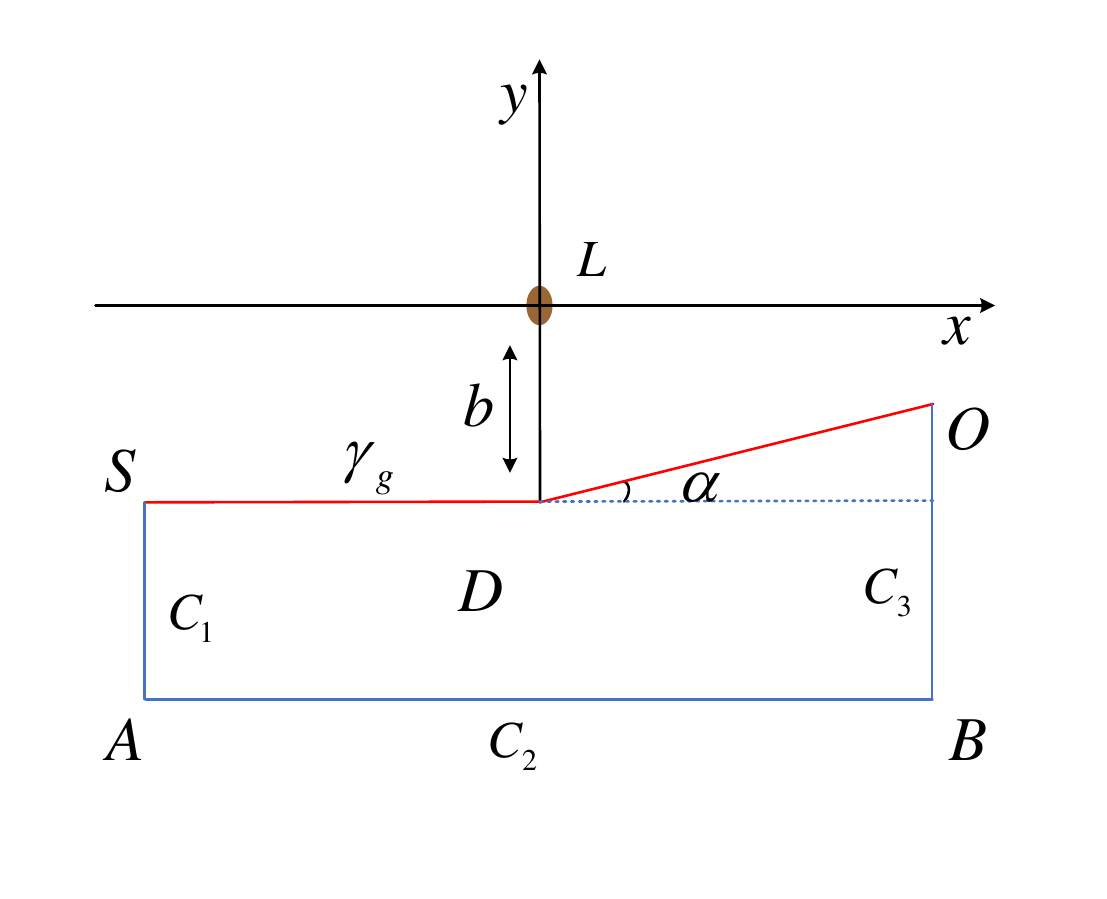}
\caption{The region $D\subset(M,\hat{g}_{ij})$ with boundary $\partial D=\gamma_g \bigcup C_i(i=1,2,3)$. Particle ray $\gamma_g$ is a spatial curve and $ C_i$ are three curves defined by $C_1:~x=-R$,~$C_2:~y=-R$ and $C_3:~x=R$ with the constant $R>0$. As $R\rightarrow \infty$, the points of intersection S, A, B, and O are in the asymptotically Euclidean region, where $S$ and $O$ denote the particle source and the observer, respectively. $L$ is the lens, $b$ is the impact parameter and $\alpha$ is the deflection angle.}\label{Figure1}
\end{figure}

Assume $M$ be a two-dimensional smooth manifold with coordinates $(x,y)$ and a Riemannian metric $\hat{g}_{ij}$. Now one can apply the GB theorem to the lens geometry in a region $D\subset(M,\hat{g}_{ij})$. For convenience,  $D$ is required to be asymptotically Euclidean and thus both the particle source $S$ and the observer $O$ are in the asymptotically Euclidean region. Let $\partial{D}=\gamma_g \bigcup C_i(i=1,2,3)$ with the particle ray $\gamma_g$ and three curves $C_i$. $\gamma_g$ is described by the impact parameter $b$, and the curves $C_i$ are defined by
\begin{eqnarray}
&& C_1:~{x=-R},\nn\\
&& C_2:~{y=-R},\nn\\
&& C_3:~{x=R},\nn
\end{eqnarray}
with the constant $R>0$. Since the lens $L$ is excluded in the domain $D$, $\chi(D)=1$. Additionally,  as $R\rightarrow \infty$, boundary curve intersections $S$,~$A$,~$B$ and $O$ are in the asymptotically Euclidean region, and thus one can have $k_g(C_{i})=0$,~$\theta_S+\theta_A+\theta_B=3\pi /2$, and $\theta_{O}=\pi/2+\alpha$ with the deflection angle $\alpha$. Then the GB theorem becomes
\begin{equation}
\lim_{R\rightarrow\infty}\left(\iint_D{K}dS-\int_{S}^{O}k_g(\gamma_g)d\sigma\right)+\left(\frac{3\pi}{2}+\frac{\pi}{2}+\alpha \right)=2\pi.\\
\end{equation}
Thus, the gravitational deflection angle can be written as
\begin{equation}
\label{GB-angle}
\alpha=\lim_{R\rightarrow\infty}\left(-\iint_D{K}dS+\int_{S}^{O}k_g(\gamma_g)d\sigma\right),\\
\end{equation}
as shown in Fig.~\ref{Figure1}.

\subsection{The equivalence between the Gibbons-Werner method and geodesics method}
In the discussion above, the Riemannian space $(M,\hat{g}_{ij})$ is somewhat arbitrary, which is asymptotically Euclidean and the condition of using the GB theorem is required. In the following, three cases will be discussed to show the equivalence between the Gibbons-Werner method and the geodesics method.

\subsubsection{Case 1: $K\neq0$, and $k_g(\gamma_g)=0$}
In this case, the particle ray $\gamma_g$ is a spatial geodesic in $(M,\hat{g}_{ij})$, and Eq.~\eqref{GB-angle} becomes
\begin{eqnarray}
\label{Gauss-Bonnet-1}
\alpha=-\lim_{R\rightarrow\infty}\iint_D{K}dS.
\end{eqnarray}
Indeed, this is the original consideration of Gibbons and Werner~\cite{GW2008,Werner2012} and for convenience we shall call it the narrow Gibbons-Werner method. In fact, many studies fall into this category. For light deflection, one has $(M,\hat{g}_{ij})=(M,g_{ij}^{opt})$, where $g_{ij}^{opt}$ is the corresponding optical metric of curved spacetime. For massive particles, $(M,\hat{g}_{ij})=(M,j_{ij})$, where $j_{ij}$ is the corresponding Jacobi metric of curved spacetime. In stationary spacetime, the optical metric (or Jacobi metric) is a Randers-Finsler metric. However, in these cases one can use the osculating Riemannian metric by Werner's method~\cite{Werner2012} or use Jusufi's method to avoid the Finsler metric~\cite{Jus-massive1}.

\subsubsection{Case 2: $K\neq0$, and $k_g(\gamma_g)\neq0$}

Now, the particle ray is not geodesic in a curved space, and Eq.~\eqref{GB-angle} can be written as
\begin{equation}
\label{Gauss-Bonnet-2}
\alpha=\alpha_\mathrm{Gauss}+\alpha_\mathrm{geod},\\
\end{equation}
where
\begin{eqnarray}
&&\alpha_\mathrm{Gauss}=-\lim_{R\rightarrow\infty}\iint_D{K}dS, \nn\\
&&\alpha_\mathrm{geod}=\lim_{R\rightarrow\infty}\int_{S}^{O}k_g(\gamma_g)d\sigma.\nn
\end{eqnarray}
In Refs.~\cite{IOA2017-2,IOA2018,IOA2019}, Ono \textit{et al.} considered the so-called generalized optical metric space as the lens background, and used Eq.~\eqref{Gauss-Bonnet-2} to study the deflection angle of light in stationary spacetimes.

\subsubsection{Case 3: $K=0$, and $k_g(\gamma_g)\neq0$}
In this case, we assume that $M$ is Euclidean space, and Eq.~\eqref{GB-angle} arrives at
\begin{equation}
\label{Gauss-Bonnet-3}
\alpha=\lim_{R\rightarrow\infty}\int_{S}^{O}k_g(\gamma_g)d\sigma.
\end{equation}
To our best knowledge, Eq.~\eqref{Gauss-Bonnet-3} has not been considered yet, and next it will be proved that this result is the same with the expression in the geodesics method.

The line element of a three-dimensional Euclidean space is
\begin{equation}
dl^2=dx^2+dy^2+dz^2,
\end{equation}
and a unit vector normal to the $x-y$ plane is $\pmb {n}=(0,0,1)$. The particle ray $\gamma_g$ can be denoted by $y=y(x)$, and one can define its unit tangent vector as
\begin{equation}
\pmb{T}=\frac{1}{\sqrt{1+y'^2}}\left(1,y',0\right),
\end{equation}
where $'$ denotes derivative with respect to $x$.
Therefore,
\begin{equation}
\dot{\pmb{T}}\equiv\frac{d\pmb T}{dl}=\frac{y''}{\left(1+y'^2\right)^2}\left(-y',1,0\right),
\end{equation}
and one can obtain the geodesic curvature of $\gamma_g$ in the $x-y$ plane as follows~\cite{Carmo1976}:
\begin{eqnarray}
k_g(\gamma_g)\equiv \dot{\pmb{T}}\cdot(\pmb {n}\times \pmb {T} )=\frac{y''}{\left(1+y'^2\right)^{3/2}}.
\end{eqnarray}
Then, one can calculate the deflection angle by
\begin{eqnarray}
\alpha&=&\lim_{R\rightarrow\infty}\int_{S}^{O}k_g(\gamma_g)dl \nn \\
&=&\lim_{R\rightarrow\infty}\int_{S}^{O}\frac{y''}{\left(1+y'^2\right)}dx \nn \\
&=&\left[\arctan{\left(\frac{dy}{dx}\right)}\right]\mid_{x\rightarrow -\infty}^{x\rightarrow \infty},
\end{eqnarray}
which is nothing but the formula of calculating deflection angle with geodesics method in Refs.~\cite{He2016,He2017a,He2017b}.

In short, the geodesics method just corresponds to special cases for the Gibbons-Werner method, where the GB theorem is used to Euclidean space. In other words, the geodesics method categorizes the deflection angle into the influence of geodesic curvature of particles moving in Euclidean space. Therefore, the geodesics method also has geometric meaning from the perspective of curvature.

\section{An example: the deflection of light in Kerr-Newman spacetime}\label{exam}

For the second-order post-Minkowskian approximation, the components of the metric of the Kerr-Newman spacetime in the harmonic coordinates $(t,x,y,z)$ can be written as~\cite{Lin2014,Yang2019}
\begin{eqnarray}
&&\nn g_{00}=-1+\frac{2 m}{r}-\frac{2m^2+q^2}{r^2}+\mathcal{O}(\varepsilon^3),\\
&&\nn g_{0i}=\zeta^i+\mathcal{O}(\varepsilon^3),\\
&& g_{ij}=\left(1+\frac{2m}{r}+\frac{m^2}{r^2}\right)\delta _{ij}+\frac{(m^2-q^2)x^i x^j}{r^4}+\mathcal{O}(\varepsilon^3),\nn\\
\end{eqnarray}
where~$m$ and~$q$ are the mass and electric charge of the Kerr-Newman black hole, respectively. ${\pmb x}=(x,y,z)$, $r=\sqrt{x^2+y^2+z^2}$ and $\zeta^i$ is the $i$th component of the gravitational vector potential $\pmb{\zeta}\equiv \frac{2ma}{r^3}(y,-x, 0)$, where $a$ is the angular momentum per unit mass. $\delta_{ij}$~is the Kronecker symbol and the expanding parameter $\varepsilon$ represents the black hole parameters $m$, $a$ or $q$. The above metric is expanded as the power series of the parameters $m$, $a$ and $q$, and $\mathcal{O}(\varepsilon^3)$ is the series with order greater than $2$, such as $m^3,a^3,q^3,m^2a,ma^2,...$.

For stationary spacetime, its optical geometry defined by the Randers-Finsler metric takes the form~\cite{Werner2012,Chern2002}
\begin{eqnarray}
&& F(x^i ,dx^i)=dt=\sqrt{\hat{\alpha}_{ij}{dx^i dx^j}}+\beta_i{dx^i},
\end{eqnarray}
where $\hat{\alpha}_{ij}$ is a Riemannian metric and $\beta_i$ is a one-form satisfying $\hat{\alpha}^{ij}\beta_i \beta_j < 1 $. Consider a null curve in the Kerr-Newman spacetime, $ds^2=0 $, and one can find a Randers-Finsler metric,
\begin{eqnarray}
\label{abcd}
\nn{\hat{\alpha}_{ij}}&=&\left[1+\frac{4m}{r}+\frac{7m^2-q^2}{r^2}\right]\delta_{ij} \\ &&+\frac{\left(m^2-q^2\right)x^ix^j}{r^4}H_{ij}+\mathcal{O}(\varepsilon^3),\nn\\
{\beta_{i}dX^i}&=&\frac{2m a\left(ydx-xdy\right)}{r^3}+\mathcal{O}(\varepsilon^3),
\end{eqnarray}
where
\begin{eqnarray}
&&\nn{H_{11}}=2xy,\ \ \ \ \ \nn{H_{12}}=H_{21}=y^2-x^2,\\
&&\nn{H_{22}}=-2xy,  \ \ \nn{H_{13}}=H_{31}=y z,\\
&&\nn{H_{33}}=0,\ \ \ \ \ \ \ \ \ \nn{H_{23}}=H_{32}=x z.
\end{eqnarray}

\subsection{Werner's method: $K\neq0$, $k_g(\gamma_g)=0$}
In this subsection, we will apply Werner's method~\cite{Werner2012} to calculate the gravitational deflection angle of light. The light ray is geodesic in Randers-Finsler space, and therefore, Eq.~\eqref{Gauss-Bonnet-1} can be considered. To simplify, one can study the null geodesic in the equatorial plane. Chose $z=0$ as the equatorial plane, and one can find the Kerr-Newman-Randers black hole optical metric as follows
\begin{eqnarray}
F\left(x^i,\frac{dx^i}{dt}\right)=\sqrt{\hat{\alpha}_{ij}{\frac{dx^i}{dt}\frac{dx^j}{dt}}}+\beta_i\frac{dx^i}{dt},
\end{eqnarray}
where $\hat{\alpha}_{ij}$ and $\beta_i$ are the same as those in Eq.~\eqref{abcd} except that $i$ and $j$ only run in $\{1,2\}$ here.

The Randers-Finsler metric is characterized by the Hessian~\cite{Chern2002,Werner2012}
\begin{eqnarray}
\label{Hessian}
g_{ij}(x,\pmb v)=\frac{1}{2}\frac{\partial^2F^2(x,\pmb v)}{\partial{v^i}\partial{v^j}},
\end{eqnarray}
where $x\in M$, and $\pmb v \in T_xM$ with $T_xM$ the tangent space at a given point. In order to obtain a Remannian metric $ \pmb{\bar {g} }$, one can choose a smooth nonzero vector field $ \pmb V $ over $ M $ that contains the tangent vectors along the geodesic $ \gamma_F $ such that $\pmb{V}(\gamma_F) = \pmb v $, defining
\begin{eqnarray}
\label{gijbar}
\bar  g_{ij}(x)=g_{ij}(x,\pmb{V}(x)).
\end{eqnarray}

In this construction, we can obtain a crucial result that the geodesic $ \gamma_F $ of $ (M,F) $ is also a geodesic $ \gamma_{\bar {g}} $ of $ (M,\bar {g}) $, i.e., $\gamma_F =  \gamma_{\bar {g}} $ ~\cite{Werner2012}.

Following Werner~\cite{Werner2012}, the osculating Riemannian manifold $(M,\bar{g}_{ij})$ can be used to calculate the gravitational defection angle of light. Near the undeflected light rays $y=-b$~\cite{He2016,He2017a}, one can choose the vector field as
\begin{eqnarray}
\label{vector field}
&& V^x=\frac{dx}{dt}=1+\mathcal{O}(\varepsilon),\nn\\
&& V^y=\frac{dy}{dt}=0+\mathcal{O}(\varepsilon).
\end{eqnarray}
Using Eqs.~\eqref{Hessian},~\eqref{gijbar}, and~\eqref{vector field}, finally the osculating Riemannian metric can be obtained as follows:
\begin{eqnarray}
\bar{g}_{xx}&=&1+\frac{4m}{r}+\frac{7m^2-q^2}{r^2}+\frac{\left(m^2-q^2\right)x^2}{r^4}\nn \\
&&+\frac{4m a y}{r^3}+\mathcal{O}(\varepsilon^3), \\
\bar{g}_{xy}&=&\bar{g}_{yx}=\frac{\left(m^2-q^2\right)x y}{r^4}-\frac{2m a y}{r^3}+\mathcal{O}(\varepsilon^3),\\
\bar{g}_{yy}&=&1+\frac{4m}{r}+\frac{7m^2-q^2}{r^2}+\frac{\left(m^2-q^2\right)y^2}{r^4}\nn\\
&&+\frac{2m a y}{r^3}+\mathcal{O}(\varepsilon^3),
\end{eqnarray}
with the determinant up to second order
\begin{eqnarray}
\label{det}
\det\bar g=1+\frac{8m}{r}+\frac{6a m y}{r^3}+\frac{31m^2-3q^2}{r^2}+\mathcal{O}(\varepsilon^3),
\end{eqnarray}
and the Gaussian curvature
\begin{eqnarray}
\label{Gauss-K-bar}
\bar K&=&\frac{\bar{R}_{xyxy}}{det{\bar{g}}}\nn\\
&=&-\frac{2m}{r^3}-\frac{3a m y(6x^2+y^2)}{r^7}\nn\\
&&+\frac{3(3m^2+q^2)}{r^4}+\mathcal{O}(\varepsilon^3).
\end{eqnarray}
In harmonic coordinates, Eq.~\eqref{Gauss-Bonnet-1} can be written as
\begin{eqnarray}
\label{Gauss-Bonnet-Werner}
\alpha=-\int_{-\infty}^{\infty}\int_{y_1(x)}^{\infty}\bar{K} \sqrt{\det{\bar{g}}}dydx.
\end{eqnarray}
Here $y_1(x)$ denotes the light ray up to first order (see the Appendix~\ref{AppendixA})
\begin{eqnarray}
\label{light ray-1}
y_1(x)=-b+\frac{2\left(x+\sqrt{b^2+x^2}\right)m}{b}+\mathcal{O}(\varepsilon^2).
\end{eqnarray}
Substituting Eqs.~\eqref{det},~\eqref{Gauss-K-bar} and~\eqref{light ray-1} into Eq.~\eqref{Gauss-Bonnet-Werner}, one can get the second-order deflection angle of light as follows:
\begin{eqnarray}
\label{def-ang-1}
\alpha=\frac{4m}{b}-\frac{4a m}{b^2}+\frac{3\pi\left(5m^2-q^2\right)}{4b^2}~+\mathcal{O}(\varepsilon^3),
\end{eqnarray}
which is consistent with the results in Ref.~\cite{He2017a}.

\subsection{The generalized optical metric method: $K\neq0$, and $k_g(\gamma_g)\neq0$~~~}  \label{GMGBT}
In this section we consider the Riemannian space ${}^{(3)}{M}$ defined by $\hat{\alpha}_{ij}$. The line element of ${}^{(3)}{M}$ is given by
\begin{eqnarray}
d\lambda^2=\hat{\alpha}_{ij}dx^idx^j~.
\end{eqnarray}
The light ray is the spatial curve in ${{}^{(3)}{M}}$ and following Fermat's principle, the motion equation of light ray is~\cite{IOA2017-2}
\begin{eqnarray}
\frac{de^i}{d\lambda}+{}^{(3)}\Gamma^i_{jk}e^je^k=\hat{\alpha}_{ij}\left(\beta_{k|j}-\beta_{j|k}\right)e^k,
\end{eqnarray}
where $e^i\equiv\frac{dx^i}{d\lambda}$,~${}^{(3)}\Gamma^i_{jk}$ denotes the Christoffel symbol associated with $\hat{\alpha}_{ij}$, and $|$ denotes the covariant derivative with $\hat{\alpha}_{ij}$. The existence of $\beta_i$ illustrates that the orbit of light is not the geodesic in ${}^{(3)}{M}$. Naturally, the contribution of geodesic curvature $k_g$ should be considered and we will use Eq.~\eqref{Gauss-Bonnet-2} to calculate the deflection angle. We focus on the motion of the light in the equatorial plane $(z=0)$. Then the geodesic curvature of curve $\gamma_g$ is given by~\cite{IOA2017-2}
\begin{eqnarray}
&&k_g(\gamma_g)=-\epsilon^{ijk} N_i \beta_{j|k},
\end{eqnarray}
where $\epsilon_{ijk}$ is the Levi-Civita tensor and $\pmb N $ is a unit normal vector for the equatorial plane. Then, choose the unit normal vector as $N_p=-\frac{1}{\sqrt{\hat{\alpha}^{zz}}}\delta_{p}^{z}$, and one can obtain
\begin{eqnarray}
k_g(\gamma_g)=\frac{1}{\sqrt{\det \hat{\alpha} \hat{\alpha}^{zz}}}\left(\beta_{x,y}-\beta_{y,x}\right),\label{Ldg1}
\end{eqnarray}
where $\epsilon^{zxy}=-\epsilon^{zyx}=1/\sqrt{\det \hat{\alpha}}$ has been used and the comma denotes the partial derivative.
With Eqs.~\eqref{abcd} and~\eqref{Ldg1}, one can have
\begin{eqnarray}
 \label{Ldg3}
k_g(\gamma_g)=-\frac{2am}{r^3}+\mathcal{O}(\varepsilon^3),
\end{eqnarray}
where the first-order light ray in Eq.~\eqref{light ray-1} has been used.

According to Eq.~\eqref{Gauss-Bonnet-2}, the deflection angle of the light can be divided into two parts. First, the Gauss curvature of $\hat{\alpha}_{ij}$ is
\begin{eqnarray}
\label{Gauss-3R}
K_{\hat\alpha}=-\frac{2m}{r^3}+\frac{3(3m^2+q^2)}{r^4}+\mathcal{O}(\varepsilon^3),
\end{eqnarray}
and one can calculate the part associated with Gauss curvature
\begin{eqnarray}
\nn{\alpha_\mathrm{Gauss}}&=&-\int_{-\infty}^{+\infty}\int_{-\infty}^{-b+\frac{2m\left(x+\sqrt{x^2+b^2}\right)}{b}}K_{\hat{\alpha}}\sqrt{\det \hat{\alpha}}~dy~dx\\
&&=\frac{4m}{b}+\frac{3\pi\left(5m^2-q^2\right)}{4b^2}+\mathcal{O}(\varepsilon^3).
\end{eqnarray}
Second, from Eqs.~\eqref{light ray-1} and~\eqref{Ldg3}, the part associated with geodesic curvature is
\begin{eqnarray}
\alpha_\mathrm{geod}&=&\lim_{R\rightarrow\infty}\int_{S}^{O}k_g(\gamma_g) d\lambda  \nn\\
&=&\int_{-\infty}^{+\infty}k_g(\gamma_g) \sqrt{\hat{\alpha}_{xx}}dx  \nn\\
&=&\int_{-\infty}^{+\infty}\left[-\frac{2a m}{\left(b^2+x^2\right)^{\frac{3}{2}}}\right]dx \nn\\
&=&-\frac{4a m}{b^2}+\mathcal{O}(\varepsilon^3).
\end{eqnarray}
Finally, the total deflection angle can be obtained as follows:
\begin{eqnarray}
\label{def-ang-2}
\alpha&=&\alpha_\mathrm{Gauss}\!+\!\alpha_\mathrm{geod}\nn\\
&=&\frac{4m}{b}\!-\!\frac{4a m}{b^2}\!+\!\frac{3\pi\left(5m^2-q^2\right)}{4b^2}\!+\mathcal{O}(\varepsilon^3),
\end{eqnarray}
which is consistent with the result in Eq.~\eqref{def-ang-1}.

\subsection{The geodesics method: $K=0$, $k_g(\gamma_g)\neq0$}
From second-order light ray in Eq.~\eqref{light-ray-2}, the following relation can be obtained
\begin{eqnarray}
\frac{dy}{dx}&=&\frac{2m \left(b-a\right)\left(x+\sqrt{b^2+x^2}\right)}{b^2\sqrt{b^2+x^2}}\nn\\
&&+\frac{3\left(5m^2-q^2\right)}{4b^2}\left(\frac{\pi}{2}+\arctan{\frac{x}{b}}+\frac{b x}{b^2+x^2}\right) \nn\\
&&-\frac{4bm^2}{\left(b^2+x^2\right)^{\frac{3}{2}}}+\frac{b\left(m^2-q^2\right)x}{2\left(b^2+x^2\right)^2}+\mathcal{O}(\varepsilon^3).
\end{eqnarray}
The deflection angle can be obtained by Eq.~\eqref{Gauss-Bonnet-3}
{\begin{eqnarray}
\label{def-ang-3}
\alpha&=&\left[\arctan{\left(\frac{dy}{dx}\right)}\right]\mid_{x\rightarrow -\infty}^{x\rightarrow \infty}\nn\\
&=&\frac{4m}{b}\!-\!\frac{4a m}{b^2}\!+\!\frac{3\pi\left(5m^2-q^2\right)}{4b^2}\!+\mathcal{O}(\varepsilon^3).
\end{eqnarray}
Certainly, this expression is the same as the result obtained by Werner's method in Eq.~\eqref{def-ang-1} and by the generalized optical metric method in Eq.~\eqref{def-ang-2}.
\section{conclusion} \label{CONCLU}
In this work, we investigate the equivalence of the Gibbons-Werner method to the geodesics method in the study of gravitational lensing. It is shown that the geodesics method can be derived with the Gibbons-Werner method for asymptotically flat spacetime. In the Gibbons-Werner procedure, one can choose the Euclidean space as the lens background and the deflection effect is completely determined by the geodesic curvature of the particle's trajectory. Thus, one can choose arbitrary asymptotically Euclidean space as the lens background and the deflection angle can be written as $\alpha=\alpha_\mathrm{Gauss}+\alpha_\mathrm{geod}$. The difference between these different background spaces is that the contribution on $\alpha_\mathrm{Gauss}$ and $\alpha_\mathrm{geod}$ is different. However, the total deflection angle is always constant. In practice, it is more convenient to use the geodesics method or the narrow Gibbons-Werner method. We can illustrate these two methods using the following formula
\begin{eqnarray}
\left[\int_{S}^{O}k_g(\gamma_g)d\sigma\right]\mid_\mathrm{Euclidean}=\left[-\iint_D{K}dS\right]\mid_\mathrm{Optical}~.\nn
\end{eqnarray}
The left side of the equation represents the geodesic method $(\alpha_\mathrm{Gauss}=0,~\alpha=\alpha_\mathrm{geod})$, while the right side represents the narrow Gibbons-Werner method $(\alpha_\mathrm{geod}=0,~\alpha=\alpha_\mathrm{Gauss})$.

As an example to show the equivalence, we calculate the second-order gravitational deflection angle of light in Kerr-Newman spacetime, for three options with the Gibbons-Werner method, in the harmonic coordinates. More, the harmonic coordinates bring a lot of simplicity and overcome the cumbersome iterative in Ref.~\cite{LHZ2019}.

\acknowledgements
This work was supported by the National Natural Science Foundation of China under Grants No.~11405136 and No.~11847307, and the Fundamental Research Funds for the Central Universities under Grant No.~2682019LK11.

\appendix
\section{Second-order light orbit} \label{AppendixA}
In this Appendix, we calculate the second-order light ray in Kerr-Newman spacetime. For the photon, the velocity $w = 1 $, and thus Eq.~(11) in the literature~\cite{He2017a} reads
\begin{eqnarray}
\frac{dy}{dp}&=&\frac{2m \left(b-a\right)x}{b^2\sqrt{b^2+x^2}}-\frac{4 b m^2}{\left(b^2+x^2\right)^{\frac{3}{2}}}+\frac{b\left(m^2-q^2\right)x}{2\left(b^2+x^2\right)^2}\nn\\
&&+\frac{3\left(5m^2-q^2\right)}{4b^2}\left(\arctan{\frac{x}{b}}+\frac{b x}{b^2+x^2}\right)+\mathcal{O}(\varepsilon^3)~,~~~~~~~~~
\end{eqnarray}
where $p $ is the affine parameter in Kerr-Newman spacetime. With the boundary conditions $\dot{y}|_{p\rightarrow{\infty}}= \dot{y}|_{x\rightarrow{\infty}}=0$~\cite{He2017a}, one can get
\begin{eqnarray}
 \frac{dy}{dp}&=&\frac{2m \left(b-a\right)\left(x+\sqrt{b^2+x^2}\right)}{b^2\sqrt{b^2+x^2}}\nn\\
&&+\frac{3\left(5m^2-q^2\right)}{4b^2}\left(\frac{\pi}{2}+\arctan{\frac{x}{b}}+\frac{b x}{b^2+x^2}\right) \nn \\
&&-\frac{4bm^2}{\left(b^2+x^2\right)^{\frac{3}{2}}}+\frac{b\left(m^2-q^2\right)x}{2\left(b^2+x^2\right)^2}+\mathcal{O}(\varepsilon^3)~.
\end{eqnarray}
Finally, with the first-order parameter transformation $dp=dx$~\cite{He2017a} and integrating $y$, one can get the second-order light ray as follows:
\begin{eqnarray}
 \label{light-ray-2}
\nn{y}&=&-b+\frac{2\left(x+\sqrt{b^2+x^2}\right)m}{b}-\frac{2a m \left(x+\sqrt{b^2+x^2}\right)}{b^2}\\
&&-\frac{m^2+3q^2}{4b}+\frac{\left(q^2-m^2\right)b} {4\left(b^2+x^2\right)}-\frac{4 x m^2}{b \sqrt{b^2+x^2}}\nn\\
&&+\frac{3\left(5m^2-q^2\right)x\left(\frac{\pi}{2}+\arctan{\frac{x}{b}}\right)}{4b^2}+\mathcal{O}(\varepsilon^3)~,
\end{eqnarray}
where we have considered the boundary conditions $y|_{p\rightarrow{\infty}}= y|_{x\rightarrow{\infty}}=-b$~\cite{He2017a}.

\end{document}